\documentclass[twocolumn,10pt]{article}
\usepackage[utf8]{inputenc}

\usepackage[
  paper=letterpaper,
  lmargin=0.8in,
  rmargin=0.8in,
  tmargin=1in,
  bmargin=1in]{geometry}

\usepackage[compact]{titlesec}
\titlespacing{\section}{0pt}{*0}{*0}
\titlespacing{\subsection}{0pt}{*0}{*0}
\titlespacing{\subsubsection}{0pt}{*0}{*0}

\usepackage[pdftex,final]{graphicx}
\usepackage[pdftex]{color}

\usepackage{authblk}

\usepackage{fancyhdr}
\usepackage[colorlinks=true,urlcolor=black,citecolor=black,linkcolor=black,final=true]{hyperref}

\usepackage[font={small,bf},textfont=md,format=hang,margin=10pt,singlelinecheck=false]{caption}

\title{Sustainable Software Ecosystems: Software Engineers, Domain Scientists,
and Engineers Collaborating for Science}
\author{Marcus D. Hanwell}
\author{Patrick O'Leary}
\author{Bob O'Bara}
\affil{Kitware, Inc., 28 Corporate Drive, Clifton Park, NY 12065 USA.\\
\url{http://www.kitware.com/}}

\begin{document}

\maketitle

\begin{abstract}

The development of scientific software is often a partnership between domain
scientists and scientific software engineers. It is especially important to
embrace these collaborations when developing advanced scientific software,
where sustainability, reproducibility, and extensibility are important. In the
ideal case, as discussed in this manuscript, this brings together teams
composed of the world's foremost scientific experts in a given field with
seasoned software developers experienced in forming highly collaborative teams
working on software to further scientific research~\cite{Ibanez_OpenScienceChapter}.

In addition to enabling scientists to perform research more effectively,
enriching the field by offering well-engineered software, sustainable
software frees
researchers from performing tasks that do not offer the rewards that their
institution values. When these software platforms are developed as collaborative
R\&D platforms, it also empowers both the team developing the software and the
wider community. We will present case studies of two DOE sponsored SBIR
projects---one in nuclear engineering that began in 2013, and another in
scanning transmission electron microscopy tomography (S/TEM) for materials.
These projects build upon the Visualization Toolkit (VTK), and ParaView, each
of which has over a decade of development history funded by multiple agencies
in collaboration with many institutions~\cite{Hanwell_WSSSPE}.

It is clear that there are examples where heroic efforts created sustainable
software, but this is clearly the exception---not the rule. Many of these
projects required significant sacrifice, and some risky bets outside of
established career paths. Their efforts should be applauded, but we must as a
community develop the necessary governance, policy, and credit mechanisms to make
sustainable, reproducible scientific software a reality. Its importance in the
sphere of scientific investigation is getting increasingly important. Many of
these points were touched upon in the the first Workshop on Sustainable
Software Ecosystems for Open Science~\cite{Katz_WSSSPE1}.

\end{abstract}

\section*{Governance}

One of the critical challenges faced when starting a new project is that of
forming the right project governance. Most projects are developed by small
teams, but if the right pieces are put in place a project can grow from humble
beginnings to effectively change the way research is done in a field. There are
many factors that contribute to forming a vibrant and sustainable community.
These include the software licenses applied, version control, mailing lists,
bug trackers, testing framework, language(s) used for development, as well as
higher level level policies such as the contribution model used, empowering
long term evangelists to move the project between funding sources, and finding
advocates to help promote the project.

Kitware, Inc. has operated as a for-profit company providing software services
and managing large software projects partnering with hundreds of institutions
for over 16 years. Over this time it has experimented with several approaches,
settling on one common approach that has worked well for several major
projects, such as VTK and ParaView~\cite{Hanwell_WSSSPE}. This approach involves
partnering with national laboratories, universities and/or other organizations
from industry where appropriate to form teams dedicated to engineering
cross-platform, open source collaborative R\&D platforms. Kitware has
traditionally focused on solutions written in C++, wrapped in Python (and
possibly other languages), although more recently this has extended to
developing more software directly in Python, and web components in HTML5,
JavaScript and WebGL.

These platforms are developed openly, with the version control, software
quality dashboards, mailing lists, and other resources all publicly hosted.
Permissive open source licenses are used, such as the 3-clause BSD license, and
the Apache 2.0 license for software, and Creative Commons licenses for data.
Software experts work hand-in-hand with partners on focused applications,
extending the underlying libraries where necessary and/or desirable. The
focused applications benefit from the significant infrastructure developed in
the software frameworks, and the frameworks benefit from additional features,
bug fixes, and enhancements added as part of the project. Kitware experts
usually work across more than one project, and will usually develop skills over
several application domains. This model has been used to develop a large number
of projects, including a new build system (CMake) that is now one of the most
widely used C/C++ build systems.

This model has been instrumental in securing funding from a large number of
funding agencies using multiple funding vehicles, along with the development of
bespoke commercial applications building upon the same frameworks.

\section*{Policies}

In order for sustainable software to become a priority the funding agencies,
universities, national labs and other institutions must adjust their policies
to reward those who develop and sustain software. The development of software
is a time consuming process, and is often not viewed in the same high regard as
publishing original research, despite the fact that it can have much wider
impacts. As the volume of data increases, and original research depends more
critically upon the data collection, analysis, and visualization of its outputs
using increasingly sophisticated software sustainable models for development of
such software becomes critical~\cite{Katz_WSSSPE1}.

Software developed in a sustainable way, across multiple institutions not only
reduces waste (through the unnecessary reinvention of the wheel), it can lead
to new research that would not otherwise have been possible---``standing on the
shoulders of software giants''. If open, permissive licenses are used we do not
unfairly restrict the application areas, and if the throw-away software
developed as part of other research projects is published under open licenses
that can be integrated into larger frameworks. If papers are able to describe
new additions to a tested software framework they can go beyond the error prone
description of algorithms using pseudocode to concrete implementations with
tests that can be verified independently.

\section*{Credit}

One of the most important aspects of sustainable software for science is that of
credit---cultivating sustainability requires an environment where credit and
incentives are in place to foster career paths and growth of long-term
projects~\cite{Ibanez_OpenScienceChapter, Hanwell_WSSSPE, Katz_WSSSPE1}. The
software development landscape is moving in the right direction---where DOIs
can be generated for software versions in Git repositories on GitHub, data on
Figshare, and other repositories. Metadata markup is being developed to more
efficiently point out who authored software, and how it should be cited.
Peer-reviewed papers addressing the development of software, and issues around
software are becoming more common.

Download statistics, service accesses, and metrics based around blogging and
social media can be used, in approaches such as altmetrics, to form a basis for
awarding credit for work, measuring wider impact, and building up a scientific
software developers portfolio. We must consider whether this is enough, and
whether funding agencies should more strongly mandate for the development of
sustainable software in science to avoid the waste in much the same way as was
done for data. Software sustainability is more complex, but also more important
as research is more computationally intensive then ever before.
Reproducibility is critical if we realistically expect to be able to build upon
the work that came before.

\section*{Case Studies}

In addition to mature, and proven projects, Kitware is at various stages of
developing new projects across a number of application domains. Here we present
two such examples that are funded by the Department of Energy (DOE) SBIR
program, the nuclear energy systems modeling is currently in Phase II, and
S/TEM materials tomography project is just a few months into Phase I of the
program.

\subsection*{Computational Analysis of Nuclear Energy Systems}

Computational analysis of nuclear energy systems is a complex multi-physics
problem. Fundamental to the solution of these problems is an appropriate
computational discrete mesh of the geometric domain of interest. A nuclear
reactor core will involve a large number of components (e.g  fuel and control
rod,  assemblies and instrumentation packages) and the simulation involves
several physical phenomena such as neutron transport, heat transfer, fluid flow,
and thermal expansion. The reactor  cores are typically arranged in either a
rectangular lattice for water-cooled reactors, or a hexagonal lattice for sodium
and gas-cooled reactors. Mesh generation for these complex geometries typically
require the coupling of various toolkits such as Meshkit from Argonne National
Laboratory~\cite{meshkit}, Cubit from Sandia National Laboratory, and
OpenCascade from OpenCascade SAS.

\begin{figure}[!hbtp]
\centering
\includegraphics[width=0.45\textwidth]{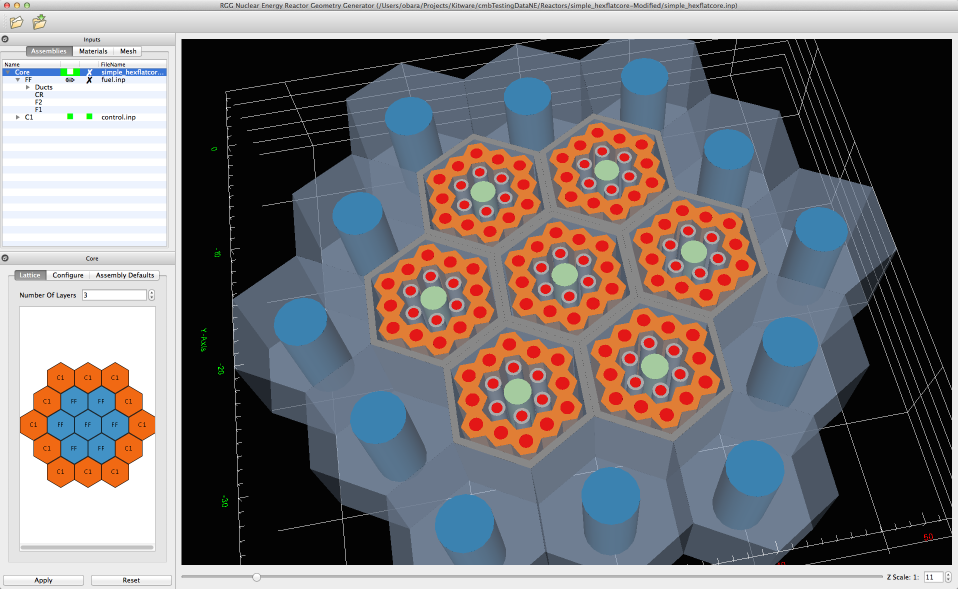}
\caption{A screenshot of the RGG application modeling a hexagonal-based
reactor core.}
\label{fig:rgg}
\end{figure}

While generation of geometry and mesh models for reactor cores can be a
difficult process, MeshKit, developed by our partners Dr. Rajeev Jain of Argonne
National Laboratory and Dr. Timothy Tautges of the University of Wisconsin
Madison, takes advantage of the various lattice structures that exist within the
core in order to reduce the complexity of model and mesh construction.
Based on a small set of parameters, geometry and mesh are constructed for a
variety of reactor core types arranged as both rectangular and hexagonal
lattices. While Meshkit is a powerful suite of tool, its target end-users are
currently the developers themselves, and not the nuclear engineers with
simulation problems.

To address this problem, an open source GUI tool called the Reactor Geometry
(and mesh) Generator (RGG) is being developed as part of a DOE Nuclear Energy
Fast Track SBIR. RGG is a cross-platform solution based upon the VTK/ParaView
and Meshkit frameworks. The first generation of the tool was developed during
the project's Phase I effort and is freely available. This version runs on
Macs, Windows, and Linux platforms. The current underlying MeshKit workflows
does require Cubit to generate the required meshes, but will be extended with
open source mesh generators early in Phase II.

During the past year, the RGG application has been presented at workshops hosted
by the DOE Nuclear Energy Advanced Modeling and Simulation (NEAMS) Program and
the Consortium for Advanced Simulation of Light Water Reactors (CASL).  The
feedback provided by the attending scientists and engineers was invaluable and
led to enhancements realized in RGG version 1.0.

In addition to contributing to the nuclear energy community, the RGG effort has
also contributed back to various open source activities. This included
identifying and correcting various deficiencies within VTK and MeskKit as well
as adding several enhancements.  In particular to MeshKit, RGG has simplified
its building process and made it possible to deploy some of MeshKit's
components to the Windows platform.  RGG has directly influenced MeshKit's
reactor modeling workflow to include the ability to share essential parametric
information between assemblies in order to generate valid meshes; resulting in a
reduction of potential end-user error. These contributions are essential to the sustainability of the
Meshkit software.

\subsection*{Atomic Resolution Materials Tomography}

Atomic resolution S/TEM tomography for materials science is a challenging area
of materials characterization with great potential. Dr. Hanwell is the
principal investigator for this project, with some background in similar
techniques, partnering with Prof. Muller and Dr. Hovden at Cornell University.
The technique has seen a number of major advances in recent years, resulting in
astounding three-dimensional images of nanosystems with atomic resolution
attainable under the right conditions. As with a number of experimental
techniques the acquisition and analysis software has not necessarily kept pace
with the hardware.

Researchers are routinely recording data at 1024x1024x1024, and sometimes even
higher resolutions. The data treatment for S/TEM tomography data can be quite
complex, with the current state-of-the-art often involving manual alignment of
the individual tilts, custom MATLAB code to perform the reconstructions, and
once the reconstructed data is ready the data size can lead to problems when
visualizing and analyzing the data. The software currently available often
lacks necessary features, and is not capable of effectively handling data of
this size. Due to the complex nature of data collection, alignment,
reconstruction and analysis it is also impossible to save the steps taken from
raw data to final reconstructed image.

A collaborative SBIR proposal was submitted recommending an open source,
cross-platform solution based on the VTK and ParaView frameworks, extending
them and creating a focused S/TEM tomography application, named TomViz, to
address these deficiencies and offer an integrated application. The Phase I
project focuses on demonstrating viability, and developing an application that
can align image slices, perform reconstruction in an extensible Python
environment built into the application, and save the complete state of analysis
using state files that can be shared, and even offered as supporting data in
publications~\cite{tomviz}.

\begin{figure}[!hbtp]
\centering
\includegraphics[width=0.45\textwidth]{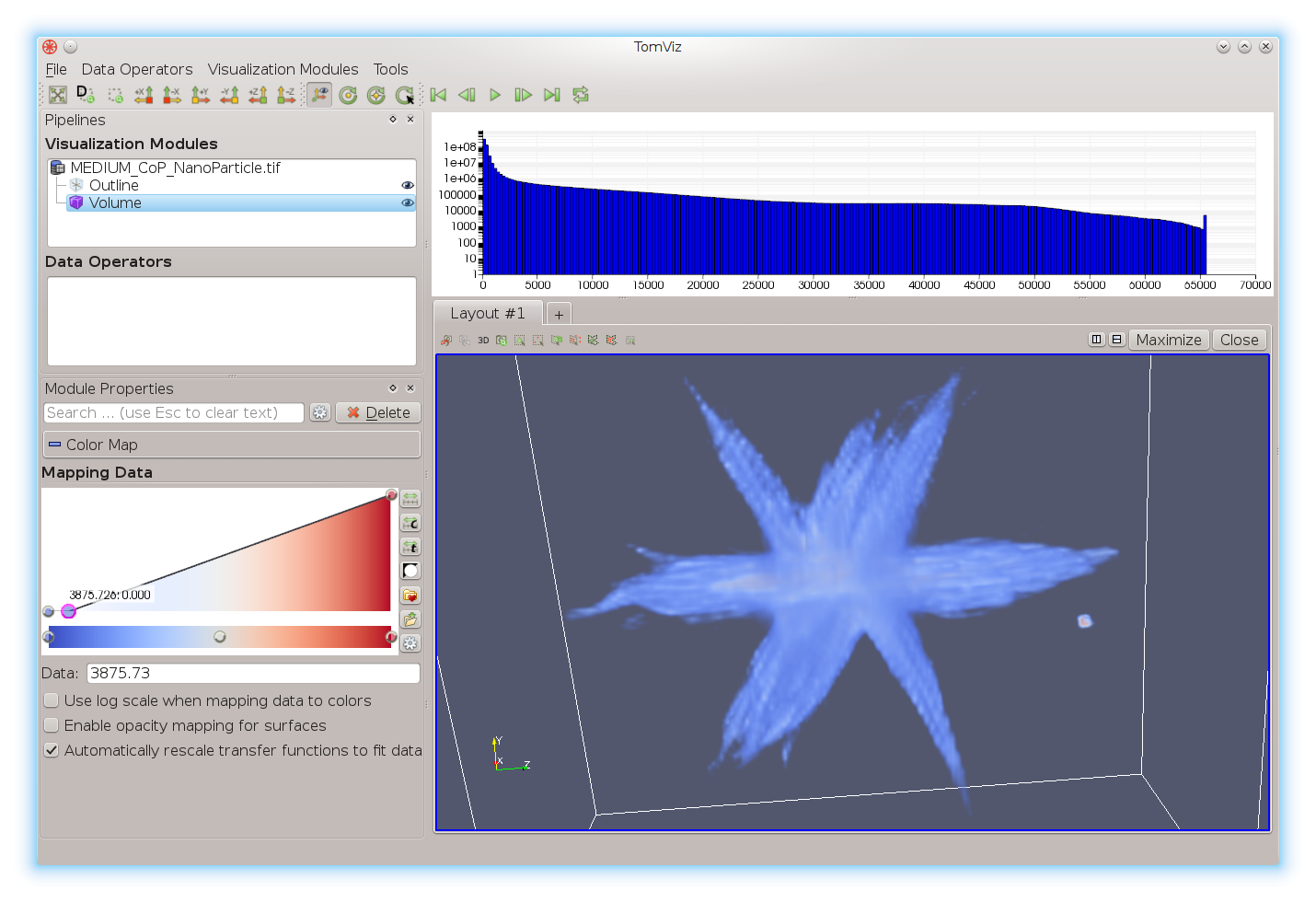}
\caption{The tomviz application showing a tomographic reconstruction of a
nanoparticle (volume rendering), with interactive histogram, and transfer
function editor.}
\label{fig:tomviz}
\end{figure}

Many of the extensions have been contributed to the VTK and ParaView projects,
making them available to the wider community. The TomViz application is
available as a binary installer for Windows, and Mac, with full source
available for other platforms. The Kitware and Cornell team will attend the
Microscopy and Microanalysis conference in August to present the first release
of the software, and has already visited Cornell to gather feedback from a
number of TEM microscopists interested in using the software in their research.
Significant advances have already been demonstrated in the analysis of
reconstructed volumes, with work on alignment and reconstruction currently
under active development in partnership with our university partners in
preparation for the scientific meeting in August. The application is shown in
Figure~\ref{fig:tomviz}, where a tomographic reconstruction of a nanoparticle
is shown.

\subsection*{Conclusions}

Software development is often seen as an afterthought when purchasing new
equipment, or an incidental output of research. We have presented some brief
case studies of early-stage projects being developed by software engineering
experts in partnership with national laboratories and universities. Without the
partnerships with software specialists researchers in these areas would not have
been able to develop software capable of tackling their research problems so
effectively, nor would they have been motivated due to current reward
mechanisms. In addition to the immediate benefits described, Kitware, as a
for-profit company, offers support services for the software, along with
consultancy services to provide new features not developed as part of the
initial project that may receive little or no interest from the research
community.

Software development is inherently a service-based industry, with the majority
of global software development being performed as a service. As a result of
abandoning the traditional license-based model Kitware has become an engine of
change, enabling the next generation of researchers to tackle some of the most
computationally challenging problems through the use of open, verified software
with a process backed by professional software developers. All source code is
available, and every step of the visualization and analysis process can be
verified, with any bugs rapidly corrected by using an established contribution
model to ensure the maximum number of people benefit. The company has a
demonstrated ability to deliver quality software, enabling open, reproducible
research that is both sustainable and through the use of permissive open source
licenses promotes shared ownership of frameworks by the wider scientific
community.

As the scientific community embraces open, reproducible scientific research
paradigms it is critical that the software infrastructure and applications keep
pace. It is not enough to simply treat software as a black-box, and expect it
to be developed because it is needed. We need to change the research landscape
to encourage teaming, take projects beyond single-person, or even single group
projects. Instead of rewarding only papers and citations we must embrace and
encourage low-friction collaboration, rewarding contributions to larger, more
established projects to avoid unnecessary reinvention of the wheel. We must
also fund and reward projects placing an emphasis on software engineering
principles, automated testing, code review, and community engagement using
software processes that support collaboration.

The computational power, and volumes of data available in all fields of
research are increasing. It is critical that we as a community embrace open,
and reproducible science, along with the role that software plays in this
ecosystem. Funding must be made available not only for novel research, but for
essential maintenance and improvement of established projects in much the same
way as central user facilities are maintained. Finally, sustainable software
must fill an important need, and be architected well enough so that it can be
used in flexible and evolving ways.

\section*{Acknowledgments}

We acknowledge DOE Office of Nuclear Energy contract DE-SC0010119 (nuclear
energy systems), and DOE Office of Science contract DE-SC0011385 (S/TEM
materials tomography) for their financial support of the projects described in
the case studies.

\subsection*{License}

This document is released under the Creative Commons Attribution 4.0 license
(CC-BY), see \url{http://creativecommons.org/licenses/by/4.0/}.

\bibliographystyle{vancouver}

\bibliography{references}

\end{document}